\documentclass[journal=jacsat,manuscript=article]{achemso}
\usepackage[version=3]{mhchem} 
\usepackage{float}

\usepackage[english]{babel}
\usepackage{array}
\usepackage{siunitx}

\usepackage[T1]{fontenc}
\usepackage[usenames,dvipsnames]{xcolor}
\usepackage{setspace}
\usepackage[compact]{titlesec}
\usepackage{hyperref}

\usepackage{epstopdf}
\usepackage{soul} 

\definecolor{cream}{RGB}{222,217,201}

\author{Alice Balbi}
\author{Andrea Rygg Aagaard}
\author{Sarai Dery Folkestad}
\author{Ida-Marie Høyvik}
\email{ida-marie.hoyvik@ntnu.no}

\affiliation[NTNU]
{Department of Chemistry, Norwegian University of Science and Technology}

\title{The role of charge resonances in the benzene dimer}

\begin{document}

\begin{abstract}
 Modern electronic-structure theory defines dispersion interactions as connected intramonomer excitations. Using this definition, dispersion contributions have been shown in literature to be large relative to other contributions at van der Waals distances for the ground state benzene dimer.
 However, are the dispersion contributions sufficient to describe  its potential energy surface? 
 In this paper, we show the importance of charge resonances for  the shape of the potential energy surface of the stacked benzene dimer. \emph{Charge resonances} is a colloquial term for the presence of ion-pair  configurations in the electronic wave function, and they represent a charge delocalization between the benzene molecules.
 We show that the ion-pair configurations, generated from connected intra- and intermonomer excitations, have a significant impact on the potential energy curves as functions of parallel displacement, as well as intramonomer separation. For parallel displacement, the energy minimum shifts approximately $\SI{2}{\angstrom}$ toward greater displacement if ion-pair configurations are not included. 
 Hence, to understand the non-covalent bonding in the benzene dimer two mechanisms must be taken into account: dispersion interaction and charge resonances.

\end{abstract}

\section{Introduction}
\label{sec:introduction}

London dispersion forces are  long-range attractive interactions, and as the separation, $R$,  increases, the dispersion energy exhibits an $R^{-6}$ decay behavior which can be shown using a low-order multipole expansion. 
However, as has been discussed by others,\cite{Jeziorski:1994aa,Stone:2013ab,Truhlar:2019aa,Herbert:2021ab} this  expansion is only valid at long-range. At van der Waals distances, charge clouds of molecules have some overlap, and it has therefore been advocated that the attractive non-covalent interaction at this length-scale should not be called dispersion.\cite{Truhlar:2019aa} Nonetheless, from correlated electronic-structure calculations, dispersion may be defined in terms of connected intramonomer excitations in the wave function.\cite{Jeziorski:1994aa,Saebo:1993aa,Hampel:1996aa,Schutz:1998aa,Azar:2012aa,Thirman:2015aa,Schneider:2016aa,Altun:2019aa} 
This definition of dispersion exhibits the expected $R^{-6}$ decay\cite{Schneider:2016aa,Hoyvik:2012ad} while being well-defined also at van der Waals distances. By well-defined, we mean that in a local orbital space representation the intramonomer separation is large enough for a clear interpretation of configurations entering the wave function.  We also note that dispersion (and other) contributions can be estimated from an energy decomposition analysis, and we refer the reader to Refs. \citenum{Jeziorski:1994aa,Azar:2012aa,Hopffgarten:2012aa,Patkowski:2020aa,Zhao:2018aa,Herbert:2021ab,Bistoni:2020aa} for good overviews of these. By exploring different energy contributions, including charge-transfer, it can be seen that dispersion is the dominant attractive force also at van der Waals distances.\cite{Schneider:2016aa,Bistoni:2020aa} However, the central question is whether dispersion contributions are sufficient to describe the correct shape of the energy surface in the van der Waals region of a dispersion dominated complex.

In this paper, we consider a prototypical example of dispersion forces\cite{Avoird}: the stacked benzene dimer. We 
investigate the shape of the potential energy curve with respect to two coordinates: an  intermonomer separation coordinate perpendicular to the benzene planes and  a parallel displacement coordinate. The  electronic wave function is  a superposition of configurations, and we directly evaluate the impact of omitting certain types of configurations, thereby circumventing the use of energy decomposition schemes. In particular,  we show that ion-pair configurations (where there is an excess negative charge on one monomer and excess positive charge on the other monomer,  e.g., M$^+$M$^-$) are crucial to describe the correct shape of the potential energy surface for the benzene dimer. Opposite ion-pair configurations enter with  equal contributions, yielding zero net charge for each monomer. The presence of ion-pair configurations in the wave function is often discussed using the term charge-transfer. However, we prefer the terms charge resonances or charge delocalization,
since we are not considering a dynamic process of charge-transfer. The presence of ion-pair configurations rather reflects that the molecules share electronic charge.

When discussing the importance of ion-pair configurations in the electronic wave function, it is imperative to consider the connection between such configurations and basis set superposition error (BSSE). BSSE is an artifact of the improved description of one monomer in the presence of the basis set on the other monomer, thereby overestimating the binding energy. Some ion-pair configurations are responsible for this type of error, and this has been pointed out in connection with local correlation models.\cite{Saebo:1993aa,Hampel:1996aa,Schutz:1998aa,Schneider:2016aa} In a coupled cluster context, BSSE is attributed to doubly ionic configurations\cite{Saebo:1993aa,Hampel:1996aa,Schutz:1998aa} (e.g., M$^{2+}$M$^{2-}$) resulting from connected double excitations. 
When investigating the effects of ion-pair configurations, a correction for BSSE must be used.
We note that while some claim that charge delocalization (charge-transfer) to mainly be a consequence of BSSE, \cite{Stone:1993aa}  others argue that such effects do not vanish in the complete basis-set limit.\cite{Schneider:2016aa}

The benzene dimer has been investigated numerous times (see e.g. Refs. \citenum{Hobza:1990aa,Hobza:1993aa,Hobza:1996aa,Tsuzuki:2002aa,Tsuzuki:2002ab,Sinnokrot:2004aa,sato:2005aa,hill:2006,prakash:2008,Pitonak:2008aa,Bludsky:2008aa,Lee:2007aa,Schneider:2016aa,herbert2020,Herbert:2021ab,Czernek:2024aa,Fanta:2025aa}) and it is reasonable to ask why the role of ion-pair configurations has been elusive. 
We believe there may be several reasons for this.
First, it is only at relatively short distances ion-pair configurations may contribute to the wave function. The Hamiltonian matrix elements which couple ion-pair configurations to neutral configurations (referred to as electronic coupling elements in the electron transfer literature) exhibit an exponential distance decay. \cite{Mikkelsen:1987aa,barbara_contemporary_1996,Subotnik:2009aa} Hence, they may only contribute to energy lowering in a limited region of the potential energy surface. 
Second, in popular energy decomposition variants such as symmetry-adapted perturbation theory,\cite{Jeziorski:1994aa,Szalewicz:2012aa,Jansen:2014aa,Patkowski:2020aa} the charge resonance (charge-transfer) energy is grouped together with induction, making it difficult to investigate the role of charge-transfer itself. Third, when charge delocalization is considered, the focus is usually on the amount of charge-transfer energy at specific geometries and relative to dispersion.\cite{Schneider:2016aa,Bistoni:2020aa} 
However, if the energy stabilization due to ion-pair configurations is large relative to the interaction energy itself, it may impact the shape of the potential energy surface significantly, 
since this contribution has an exponential distance decay (as discussed above).

In this paper, we avoid energy decomposition approaches altogether, and rather exploit the fact that we can control which type of configurations are included in the optimization of the electronic wave function. We are able to demonstrate that, contrary to common belief, there are two vital mechanisms for the attractive interaction in the benzene dimer. One mechanism (charge delocalization between the monomers) is important for the shape of the potential energy curve around van der Waals distances.  The other (dispersion) dictates the long-range behavior.

\section{Theory}
\label{sec:theory}
The molecular electronic Hamiltonian in the orbital basis is given by,
\begin{equation}
\label{eq:hamiltonian}
\hat{H} =\sum_{PQ} h_{PQ}E_{PQ}+\frac{1}{2}\sum_{PQRS}g_{PQRS}e_{PQRS}+h_\mathrm{nuc}
\end{equation}
where $E_{PQ}$ and $e_{PQRS}$ are the one- and two-electron singlet excitation operators in the second quantization formalism, respectively, and $h_{PQ}$ and $g_{PQRS}$ are the one- and two-electron integrals. See e.g., Ref. \citenum{helgaker2013molecular}, chapter 2, for details.
The coupled cluster wave function,$|\Psi_\mathrm{CC}\rangle$, is written as 
\begin{equation}
\label{eq:cc_state}
|\Psi_\mathrm{CC}\rangle =\exp(\hat{T})|\Phi\rangle
\end{equation}
where $\hat{T}$ is the cluster  operator which generates excitations out of the reference state, $|\Phi\rangle$. The reference state may be chosen to be the Hartree--Fock determinant. The excitation rank of $\hat{T}$ depends on the  chosen model. For closed-shell coupled cluster singles and doubles (CCSD)\cite{Purvis:1981aa}, the operator, in the orbital basis, is given by
\begin{equation}
\label{eq:cluster_canonical}
\hat{T} = \sum_{IA} t_I^A E_{AI} + \frac{1}{2}\sum_{IJAB} t_{IJ}^{AB} E_{AI}E_{BJ},
\end{equation}
where $t_I^A$ and $t_{IJ}^{AB}$ are the singles and doubles cluster amplitudes, respectively, and $E_{AI}$ is a singlet excitation operator,  exciting an electron out of orbital $I$ and into orbital $A$. Here,  $I$ and $J$ denote occupied Hartree--Fock orbitals and $A$ and $B$ denote unoccupied (virtual) Hartree--Fock orbitals.  When $\exp(\hat{T})$ acts on the reference determinant, as in eq. \eqref{eq:cc_state}, it generates a linear combination of up to N-tuply excited determinants.

In its standard formulation, the coupled cluster wave function is expressed in a set of canonical Hartree--Fock orbitals, which are delocalized across the molecular system.  However, the Hartree--Fock state is invariant with respect to rotations among occupied orbitals and among virtual orbitals. Hence, there are infinitely many choices of orbitals which describe the exact same Hartree--Fock state, and we may choose a set of orbitals which is convenient for the problem at hand.  

In this work, we express the coupled cluster state (eq. \eqref{eq:cc_state}) in a set of spatially localized orthogonal Hartree--Fock orbitals,\cite{Hoyvik:2016aa} for the purpose of investigating the noncovalent interaction between two benzene molecules, referred to as  $\mathcal{A}$ and $\mathcal{B}$ for simplicity. Hence, from spatially localizing the canonical set of occupied  and virtual orbitals, we obtain occupied and virtual orbitals which are centered either on  $\mathcal{A}$ or $\mathcal{B}$. Since there is no covalent bond between $\mathcal{A}$ and $\mathcal{B}$, there will be no orbitals which are shared by the two, but we note that due to the orthogonality requirement, small tail components of orbitals centered on $\mathcal{A}$ will be present on $\mathcal{B}$, and vice versa. 

The full set of occupied and virtual Hartree--Fock orbitals ($\{\phi_I\},\{\phi_A\}$) is spanned by 
\begin{equation}
\begin{split}
&\{\phi_p\}=\{\phi_i\},\{\phi_a\}\quad \text{centered on monomer $\mathcal{A}$}\\
&\{\phi_{\bar p}\}=\{\phi_{\bar i}\},\{\phi_{\bar a}\}\quad \text{centered on monomer $\mathcal{B}$}
\end{split}
\end{equation}
and we re-emphasize that the sets are mutually orthogonal. In the localized basis, the standard Hartree--Fock state of the composite system  $\mathcal{A}\mathcal{B}$ is given by
\begin{equation}
\label{eq:hf_determinant}
|\Phi\rangle = \prod_i^{N_\mathcal{A}/2} a_{i \alpha}^\dagger a_{i\beta}^\dagger\prod_{\bar i}^{N_\mathcal{B}/2} a_{\bar{i} \alpha}^\dagger a_{\bar{i}\beta}^\dagger|\text{vac}\rangle
\end{equation}
where $N_\mathcal{A}$ and $N_\mathcal{B}$ are the number of electrons on $\mathcal{A}$ and $\mathcal{B}$ as given by the Hartree--Fock orbitals (42 electrons on each benzene molecule).  The form in eq. \eqref{eq:hf_determinant} implicitly defines the Hartree--Fock state to be without any charge delocalization between the molecules. Hence, any contribution of ionic configurations are coming through the correlated part of the wave function.

\begin{figure}[H]
    \centering
    \includegraphics[width=0.8\linewidth]{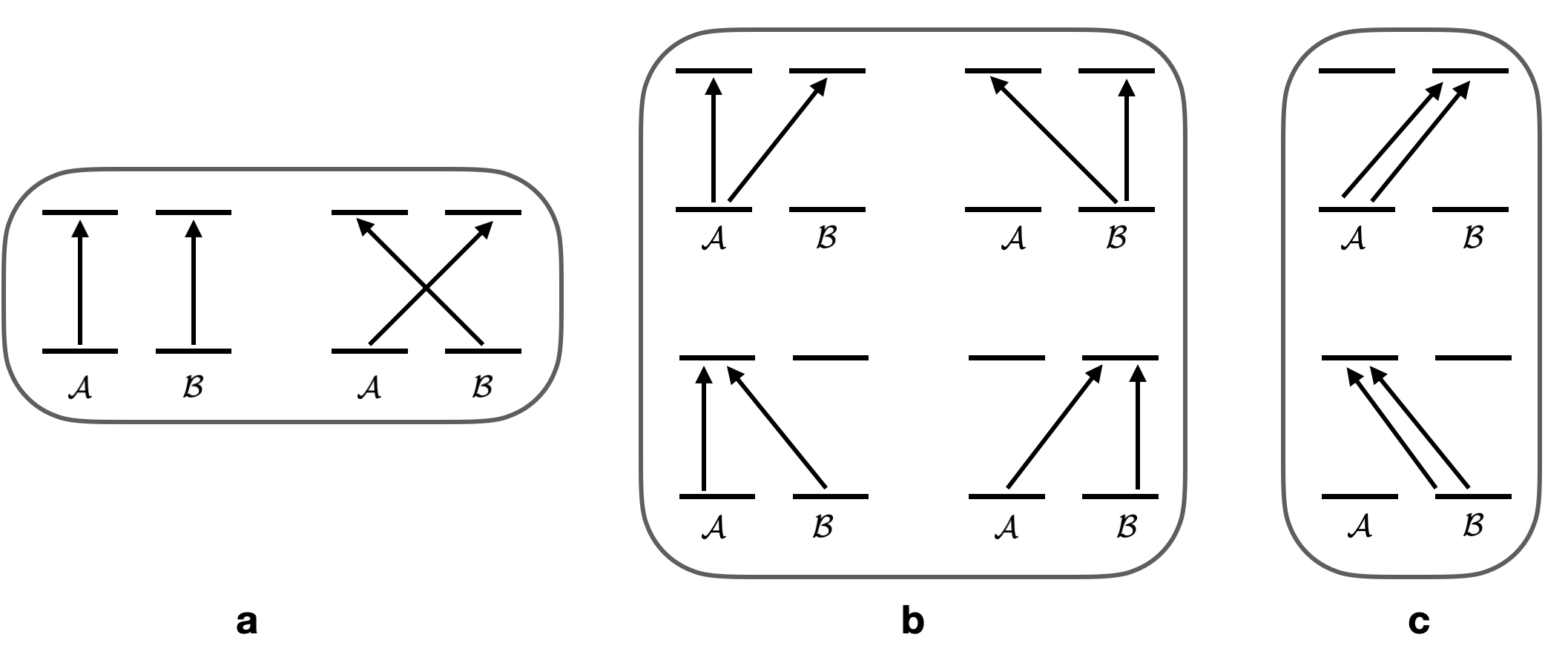}
    \caption{An illustration of connected doubles excitations in $T_\mathcal{AB}$. a: excitations which are in $\hat{T}_\mathcal{AB}^\text{neutral}$. The two processes represent dispersion. b: Processes which generate singly ionic configurations and which are in $\hat{T}_\mathcal{AB}^\text{ionic}$. c: Processes which generate doubly ionic configurations and are also in $\hat{T}_\mathcal{AB}^\text{ionic}$.}
    \label{fig:amplitudes}
\end{figure}

The cluster operator in eq. \eqref{eq:cluster_canonical} can thus be written as,
\begin{equation}
\label{eq:cluster_local}
\hat{T} = \hat{T}_\mathcal{A} + \hat{T}_\mathcal{B} + \hat{T}_\mathcal{AB}
\end{equation}
where $\hat{T}_\mathcal{A}$ contain all terms which refer only to orbitals in $\{\phi_p\}$,  $\hat{T}_\mathcal{B}$ contain all terms which refer only to orbitals in $\{\phi_{\bar p}\}$ and $\hat{T}_\mathcal{AB}$ contain terms which refer to both orbital spaces. Hence, $\hat{T}_\mathcal{AB}$ contain the terms which describes direct interaction within the correlated picture.  We will now divide the terms in  $\hat{T}_\mathcal{AB}$ into two categories; those excitations which generate neutral configurations ($\hat{T}_\mathcal{AB}^\text{neutral}$) 
and those who generate singly and doubly ionic configurations ($\hat{T}_\mathcal{AB}^\text{ionic}$).
\begin{equation}
\label{eq:cluster_ab}
\hat{T}_\mathcal{AB} =  \hat{T}_\mathcal{AB}^\text{neutral}+\hat{T}_\mathcal{AB}^\text{ionic}
\end{equation}
In Fig. \ref{fig:amplitudes} the connected doubles excitations inside $\hat{T}_\mathcal{AB}$ are illustrated. In Fig. \ref{fig:amplitudes}a excitations which belong to $\hat{T}_\mathcal{AB}^\text{neutral}$ are shown. They represent what is commonly referred to\cite{Schneider:2016aa,Bistoni:2020aa,Martin:2022aa} as (genuine) dispersion and exchange-dispersion. In Fig. \ref{fig:amplitudes}b connected double excitations resulting in ion-pair configurations (a net transfer of  one electron from one benzene to the other, $\mathcal{A}^-\mathcal{B}^+$ and $\mathcal{A}^+\mathcal{B}^-$) are shown. In Fig. \ref{fig:amplitudes}c connected double excitations which generate doubly ionic ion-pair configurations ($\mathcal{A}^{2-}\mathcal{B}^{2+}$ and $\mathcal{A}^{2+}\mathcal{B}^{2-}$) are illustrated.   Excitations in \ref{fig:amplitudes}b-c belong to $\hat{T}_\mathcal{AB}^\text{ionic}$. 
Furthermore, $\hat{T}_\mathcal{AB}^\text{ionic}$ contains single excitations between $\mathcal{A}$ and $\mathcal{B}$,
whereas  $\hat{T}_\mathcal{AB}^\text{neutral}$ only has double excitations (since all neutral single excitations belongs to $\hat{T}_\mathcal{A}$ and $\hat{T}_\mathcal{B}$).  We note that the presence of ion-pair configurations does not mean that the dimer itself has ionic character, since opposite ion-pair configurations occur with equal weight in the wave function. 

From eq. \eqref{eq:cluster_local} and \eqref{eq:cluster_ab}, we may define a version of CCSD where only neutral configurations enter, i.e., where $\hat{T}_\mathcal{AB}^\text{ionic}$ is omitted,  and which we refer to as \emph{charge-localized CCSD} (cl-CCSD). The cluster operator for cl-CCSD is given by,
\begin{equation}
\label{eq:cluster_cl}
\hat{T}_\mathrm{cl} = \hat{T}_\mathcal{A} + \hat{T}_\mathcal{B} +\hat{T}_\mathcal{AB}^\text{neutral}
\end{equation}
At this point it is also worth noting the connection of the cl-CCSD model presented here to models developed to  quantify charge-transfer and charge delocalization, such as the active space decomposition method\cite{Parker:2013aa,Parker:2014ab} and the charge-localized determinant framework for configuration interaction models.\cite{folkestad:2025}

\section{Methodology}
\label{sec:methodology}

We will present results for both CCSD and cl-CCSD (see eq. \eqref{eq:cluster_cl}) using aug-cc-pVDZ\cite{Dunning:1989aa} basis set on carbon atoms and cc-pVDZ on hydrogen atoms. We  note that while the  CCSD/aug-cc-pVDZ model is not sufficient for highly accurate  interaction energies, the approach captures the essence of the benzene dimer interactions. It thus serves as a computationally efficient model for comparing CCSD and cl-CCSD. In cl-CCSD, all terms in $\hat{T}_\mathcal{AB}^\text{ionic}$ of eq. \eqref{eq:cluster_ab} are set to zero. The coupled cluster equations are solved directly in the local orbital basis, and in cl-CCSD calculations the ion-pair amplitudes are kept zero throughout solving the amplitude equations.   All interaction energies presented are corrected for BSSE using the counterpoise correction of Boys and Bernardi,\cite{Boys:1970aa}  i.e., at each geometry the counterpoise corrected interaction energy is computed. We note that an alternative way of correcting for BSSE,  would be to set doubly ionic amplitudes to zero in CCSD, as these are identified to be responsible for BSSE.\cite{Saebo:1993aa} However, we prefer to keep CCSD as commonly defined and rather use a standard approach for BSSE correction.

\section{Results}
\label{sec:results}

In this section we investigate the role ion-pair configurations play for the potential energy interaction energy curves with respect to  the horizontal displacement ($d$) and vertical intermolecular separation  ($R$)  of the parallel displaced benzene dimer relative to the equilibrium geometry. The monomers are kept frozen upon changing $d$ and $R$.  The starting geometry was taken from Table S3 in Ref. \citenum{Sinnokrot:2004aa},  where it was optimized at the estimated CCSD(T) and modified aug-cc-pVQZ  level of theory (see Ref. \citenum{Sinnokrot:2004aa} for details). 
 The local orbital space coupled cluster code is implemented in a development version of the $e^\mathcal{T}$ program.\cite{Folkestad:2020ac}
 Orbital localization is performed using the Foster-Boys localization function\cite{foster_canonical_1960,edmiston_localized_1963}, using the trust-region orbital localization algorithm\cite{Hoyvik:2012ab} as implemented in $e^\mathcal{T}$.\cite{Folkestad:2022aa}

\subsection{The energy as a function of horizontal displacement, $d$}
\label{sec:displacement}
\begin{figure}[H]
    \centering
    \includegraphics[width=0.7\linewidth]{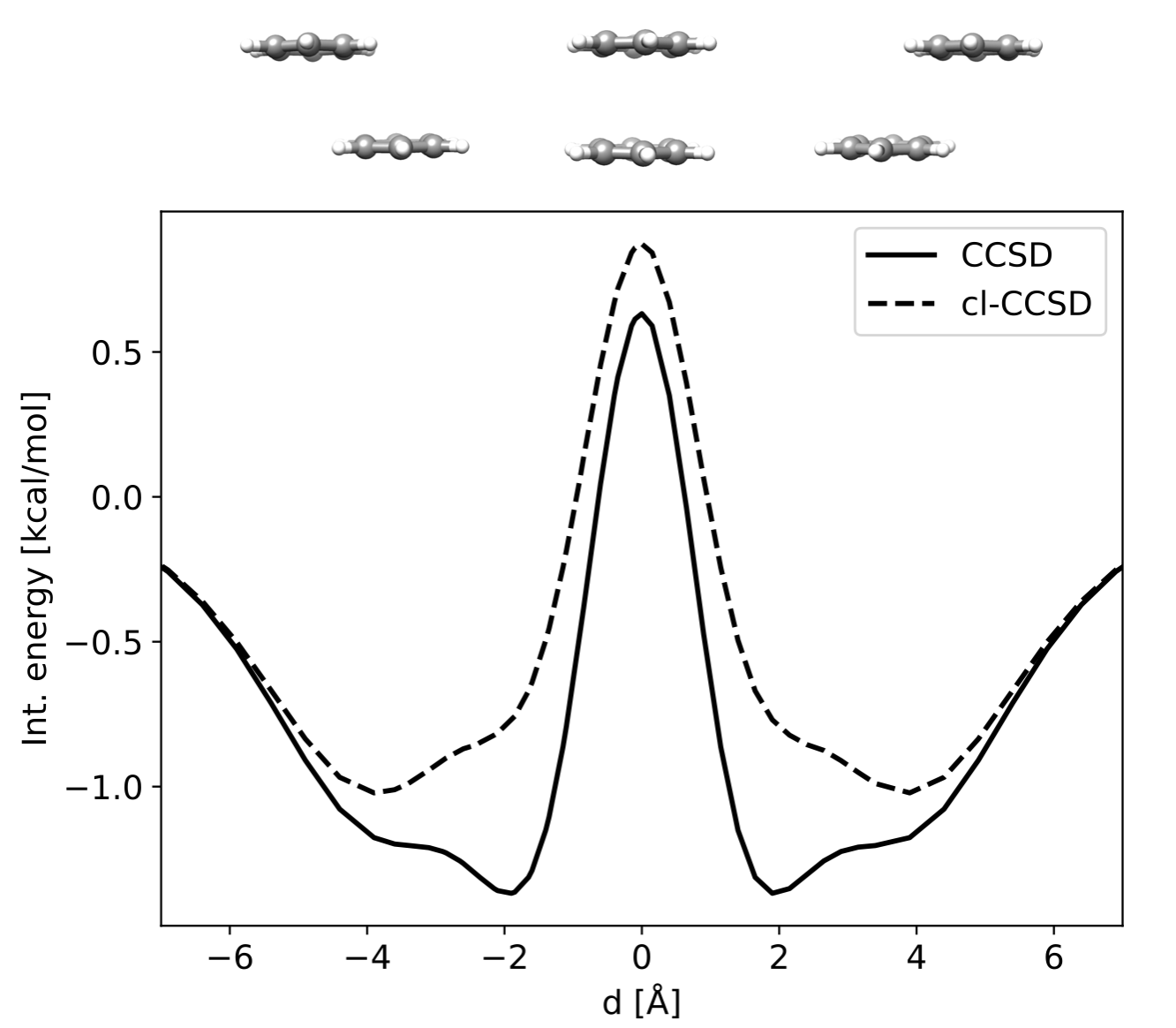}
    \caption{The interaction energy computed using counterpoise corrected CCSD and cl-CCSD  as a function of horizontal displacement, $d$, on an interval symmetric around the sandwiched ($d=0$) geometry. The calculations use aug-cc-pVDZ on the carbon atoms and cc-pVDZ on the hydrogen atoms. 
    The coordinate is illustrated by the dimer geometries shown above the plot, made using UCSF Chimera.\cite{Pettersen:2004aa}}
    \label{fig:benzene}
\end{figure}

 In Figure \ref{fig:benzene} we have  plotted the counterpoise corrected CCSD and  cl-CCSD interaction energies of the benzene dimer as a function of horizontal displacement, $d$. The intermolecular separation, $R$, is kept fixed at the equilibrium separation in the geometry taken from Ref. \citenum{Sinnokrot:2004aa}, i.e., 3.6 Å.

In the region between 
$d=0$ Å and $d=4.0$ Å, we can see qualitative differences between CCSD and cl-CCSD. At  $d=0$ Å, both curves 
are repulsive, as expected.
As $d$ increases, we see that the CCSD interaction energy  goes down to a minimum around $d= 1.9$ Å. In contrast, the interaction energy minimum for cl-CCSD is located at $3.9$ Å, i.e., without ion-pair configurations, the minimum is shifted by 2.0 Å toward greater displacement. Hence, the small contributions (relative to the magnitude of other contributions)  from ion-pair configurations  become decisive for the shape of the curve, because they are large relative to the interaction energy itself and they change rapidly  as a function of $d$. The minimum interaction energy found by  cl-CCSD is -1.02 kcal/mol, versus -1.37 kcal/mol for CCSD. Hence, in addition to moving the minimum by 2.0 Å, the interaction is not as strong without the ion-pair configurations. We note that  we computed interaction energies on a grid in $d$, we did not do geometry optimizations, so these are the minima on the grid in $d$ (see list of structures in Ref. \citenum{geometries}).

Beyond displacements of approximately $d>4.5$ Å, CCSD and  cl-CCSD become equivalent. Hence, the contributions from ion-pair configurations become irrelevant for larger displacements, as expected given that they are highly distance dependent. The Hamiltonian elements that couple neutral an ion-pair configurations are well-known from electron transfer literature, where exponential decay with distance is seen.\cite{Mikkelsen:1987aa,barbara_contemporary_1996,Subotnik:2009aa}

\subsection{The energy as a function of intermolecular separation, $R$}

\begin{figure}[H]
    \centering
    \includegraphics[width=0.75\linewidth]{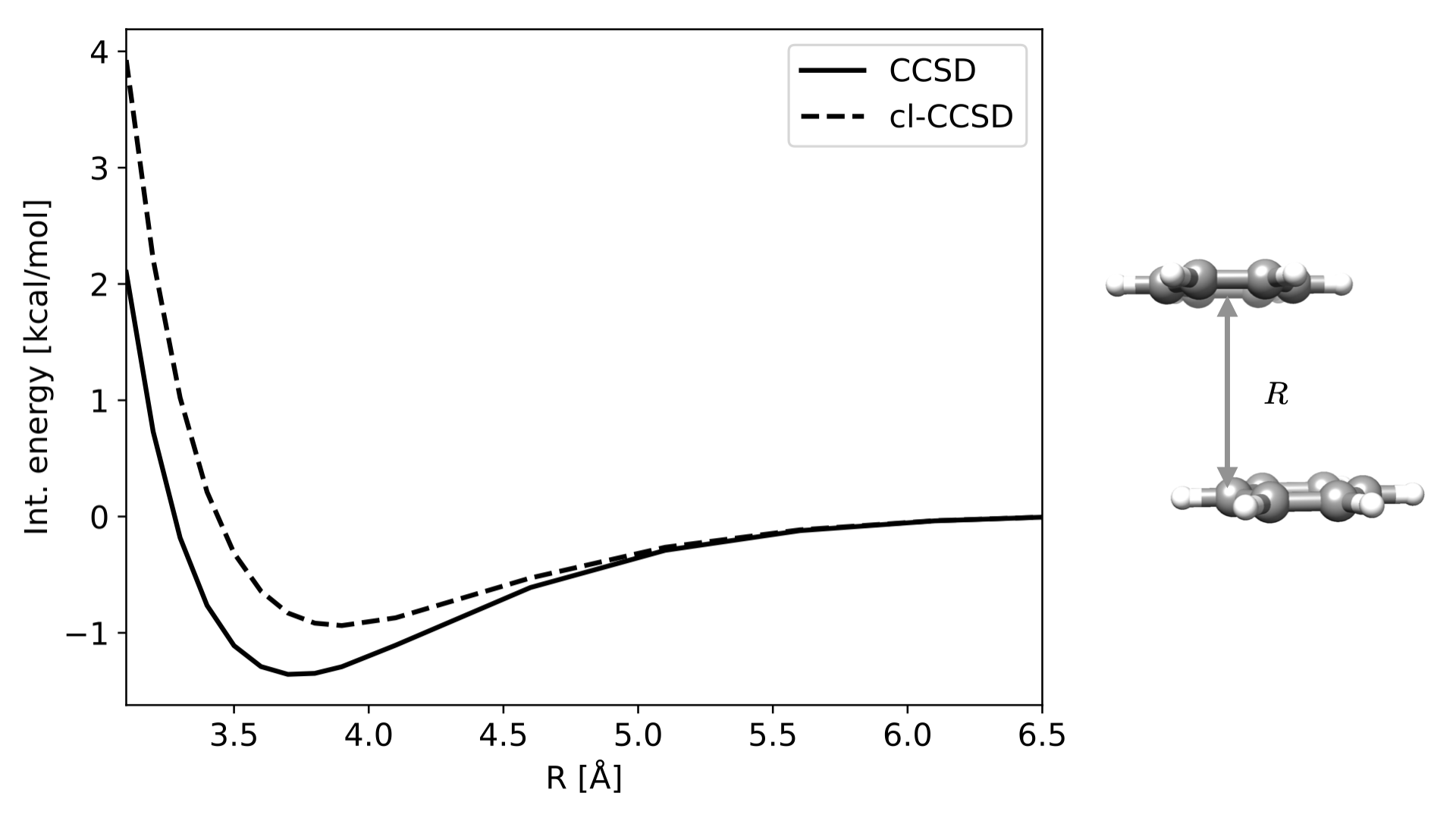}
    \caption{CCSD  and cl-CCSD  counterpoise corrected interaction energies as a function of intermolecular separation, $R$. The displacement coordinate is kept fixed at $d=1.6$ Å. The calculations use aug-cc-pVDZ on the carbon atoms and cc-pVDZ on the hydrogen atoms.  }
    \label{fig:benzene_R}
\end{figure}

In Fig. \ref{fig:benzene_R} we have plotted the counterpoise corrected CCSD and  cl-CCSD interaction energies of the benzene dimer as a function of intermolecular separation, $R$, while the displacement coordinate is kept fixed at $d=1.6$ Å (corresponding to the equilibrium displacement given in Ref. \citenum{Sinnokrot:2004aa}). 

We first note that both CCSD and cl-CCSD provide binding curves, and that they give equivalent results beyond approximately $R=4.5$ Å, where the interaction energy tends toward zero. The equivalence at larger $R$ has the same justification as discussed for the displacement coordinate in Section \ref{sec:displacement}, namely the rapid decay of elements which couple neutral and ion-pair configurations.

Considering shorter distances, the lowest energy found on our grid of $R$  (see list of structures in Ref. \citenum{geometries}) is $-1.4$ kcal/mol for CCSD and $-0.9$ kcal/mol for cl-CCSD. The minimum for CCSD is located at approximately $R=3.7$ Å, while the minimum  on the cl-CCSD curve is located at $R=3.8$ Å. Hence, there is only a small change in equilibrium distance for intermolecular separation, but cl-CCSD capture only about 64\% of the interaction energy at this  level of theory. Hence, the ion-pair configurations which are omitted in cl-CCSD, are important for stabilizing the dimer interaction.

\section{Summary and concluding remarks}
\label{sec:summary}
The results presented in this paper, show that
charge resonances (i.e., the presence of ion-pair configurations in the wave function) are necessary to describe the non-covalent attraction in the ground-state benzene dimer.
Using only neutral configurations, responsible for dispersion interactions,  provides qualitatively  wrong results for the potential energy curve of the benzene dimer. The presence of ion-pair configurations indicates a charge delocalization between the two benzene molecules, which we show to be crucial. Opposite ion-pair configurations enter with equal weights in the superposition, as there is naturally zero net charge of the monomers in the dimer. 
Examples of other terms used in literature to describe what we here refer to as charge resonance and charge delocalization are \emph{charge-transfer} or \emph{fractional covalency}. Until now, however, their importance for the description of the non-covalent bonding in the ground state benzene dimer has not been shown.

To demonstrate the importance of  ion-pair configurations, 
 we present CCSD results together with a modified form of CCSD where all amplitudes which represent ion-pair
configurations are set to zero. We refer to the latter as charge-localized (cl-)CCSD. Doubly ionic configurations in the wave functions have previously been identified as a manifestation of BSSE in interaction energy calculations, leading to severely overestimated non-covalent bonding. Since doubly ionic configurations are not present in cl-CCSD, the total energy curves for CCSD and cl-CCSD cannot be directly compared. Relative to cl-CCSD, the CCSD results will have significant BSSE. Here, we therefore presented counterpoise corrected interaction energies for both models. For a displacement coordinate, the minimum of the potential energy surface shifts $\SI{2}{\angstrom}$ if ion-pair configurations are not included. For the intermolecular separation, a third of the interaction energy is lost without these configurations while the minimum is shifted only slightly (by $\SI{0.1}{\angstrom}$). In general,  the ion-pair configurations have a large contribution to the energy at van der Waals distances, and at the same time these energy contributions vary rapidly with changes in displacement and intermolecular separation. Hence, they have a large impact on the shape of the potential energy curve.
It should therefore be recognized that  dispersion is only one of two important mechanisms that govern the non-covalent attraction of the benzene dimer.

\section*{Author contributions}

Alice Balbi: Writing--Review and editing (equal), Software--project specific (lead), Computations (lead), Formal analysis (equal). Andrea Rygg Aagard: Computations (supporting), Writing--Review and editing (equal). Sarai Dery Folkestad: Software--coupled cluster code (lead), Computations (supporting), Formal analysis (supporting), Writing--original draft (supporting), Writing--Review and editing (equal), Conceptualization (supporting). Ida-Marie Høyvik: Conceptualization (lead), Funding Acquisition (lead),  Formal analysis (equal), Writing--original draft (lead), Writing--Review and editing (equal), Computations (supporting).

\section*{Data availability}

All geometries used for producing the numbers are provided in Ref.~\citenum{geometries}.

\section*{Acknowledgments}

A.B and I-M.H. acknowledge funding from the Research Council of Norway through FRINATEK project 325574. I-M.H acknowledge funding from the European Union (ERC, OpenQuantum, 101170817). Views and opinions expressed are however those of the author(s) only and do not necessarily reflect those of the European Union or the European Research Council. Neither the European Union nor the granting authority can be held responsible for them.

\appendix

\section{The CCSD and charge-localized CCSD energies}

The standard CCSD energy is given by
\begin{equation}
E_\mathrm{CCSD}=E_\text{HF} +\sum_{IJAB}(t_{IJ}^{AB}+t_I^At_J^B)L_{IAJB}
\end{equation}
where summation over $I$ and $J$ are over all occupied orbitals on $\mathcal{A}$ and $\mathcal{B}$ ($\{\phi_i\}$ and $\{\phi_{\bar i}\}$). $L_{IAJB}$ is standard notation for two times coulomb minus exchange, i.e., 
\begin{equation}
L_{IAJB} = 2 g_{IAJB} - g_{IBJA}
\end{equation}

The CCSD energy, when excluding all single and double ion-pair configurations (using $\hat{T}_{cl} = \hat{T}_\mathcal{A} + \hat{T}_\mathcal{B} + \hat{T}_\mathcal{AB}^\text{neutral}$), is given by
\begin{equation}
\label{eq:energy_ccsd_reduced}
\begin{split}
E_\mathrm{CCSD}^\text{cl} &=E_\text{HF}+ \sum_{ijab}(t_{ij}^{ab}+t_i^at_j^b)L_{iajb} + \sum_{\bar i \bar j\bar a \bar b}(t_{\bar i \bar j}^{\bar a \bar b}+t_{\bar i}^{\bar a}t_{\bar j}^{\bar b})L_{\bar i\bar a\bar j\bar b} + 2\sum_{ijab}t_{\bar ij}^{a\bar b}L_{\bar iaj \bar b}\\
&\equiv E_\text{HF}+  E_\mathrm{CCSD}(\mathcal{A}) + E_\mathrm{CCSD}(\mathcal{B}) + E_\mathrm{exch-disp}
\end{split}
\end{equation}
Where the notation $E_\mathrm{exch-disp}$ has been introduced, according to standard interpretation of the process represented by $t_{\bar ij}^{a\bar b}$, see e.g., Ref. \cite{Schneider:2016aa}.
For simplicity, we we refer to the model where ion-pair amplitudes have been set to zero as cl-CCSD. We not that it is not just in the final energy expression they are set to zero, they are kept zero throughout solving the coupled cluster amplitude equations. Further, we note that it can easily be seen from eq. \eqref{eq:energy_ccsd_reduced}, that $E_\mathrm{CCSD}^\text{cl}$ is invariant with respect orthogonal transformations among occupied on $\mathcal{A}$, among virtual on $\mathcal{A}$, and similarly for the occupied and virtual set on $\mathcal{B}$. Hence, any such rotation of the orbitals will produce exactly the same energy.

\bibliography{rsc} 
\end{document}